\documentclass[12pt,a4paper]{article}
\usepackage[dvips]{graphicx}
\usepackage{graphics}
\usepackage{graphicx}
\usepackage{amssymb}
\usepackage{amsmath}
\usepackage{amssymb}
\usepackage{epsf}
\input{epsf}
\begin{document}
\begin{titlepage}
\title{\bf Bogoliubov's Vision   and Modern Theoretical Physics}  
%\thanks{International Journal of Modern Physics,  B24,  
%(2010)} 
%
\author{
A. L. Kuzemsky \thanks {E-mail:kuzemsky@theor.jinr.ru;
http://theor.jinr.ru/\symbol{126}kuzemsky}
%}
\\
{\it Bogoliubov Laboratory of Theoretical Physics,} \\
{\it  Joint Institute for Nuclear Research,}\\
{\it 141980 Dubna, Moscow Region, Russia.}}
\date{}
\maketitle
\begin{abstract}
A brief survey of the author's works on the fundamental conceptual ideas of quantum statistical physics
developed by N. N. Bogoliubov and his school was given. The development and applications of the method  of quasiaverages  to 
quantum statistical physics and condensed matter physics  were  analyzed.
The relationship with the concepts of broken symmetry, quantum protectorate and emergence was examined, and the progress to date towards
unified understanding of complex many-particle systems was summarized. Current trends for extending and using these
ideas in quantum field theory and condensed matter physics were discussed, including microscopic theory of superfluidity and superconductivity, 
quantum theory of magnetism of complex materials, Bose-Einstein condensation, chirality of molecules, etc.

%
%%%%%%%%%%%%%%%%%%%%%%
%
\end{abstract}
\end{titlepage}
Works and ideas of N.N. Bogoliubov and his school continue to influence and vitalize the development of modern physics~\cite{nnb1,nnbc}.  In recently 
published review article by A.L. Kuzemsky~\cite{kuz10}, which is a substantially extended version of his talk on the last 
Bogoliubov's Conference~\cite{nnbc}, the 
detailed analysis of a few selected directions of researches of N.N. Bogoliubov and his school was carried out. This interdisciplinary review 
focuses on the applications of symmetry principles to quantum and statistical physics in connection with some other branches of science.  Studies 
of symmetries and the consequences of breaking them have led to deeper understanding in many areas of science. The role of symmetry in physics is 
well-known~\cite{pwa84,wil08}.  Symmetry was and still is one of the major growth areas of scientific research, where the frontiers of mathematics and physics 
collide. Symmetry has always played an important role in condensed matter physics, from fundamental formulations of basic principles to concrete 
applications.  Last decades show clearly its role and significance for fundamental physics. This was confirmed by awarding the Nobel Prize to Y. 
Nambu et al. in 2008. In fact, this Nobel Prize should be shared with N.N. Bogoliubov, whose works influenced Y. Nambu greatly.\\ 
The article [3] 
examines  the Bogoliubov's notion of quasiaverages, from the original paper, through to modern theoretical concepts and ideas of how 
to describe both the degeneracy, broken symmetry and the diversity of the energy scales in the many-particle interacting systems. Current trends 
for extending and using Bogoliubov's ideas to quantum field theory and condensed matter physics problems were discussed, including microscopic 
theory of superfluidity and superconductivity, quantum theory of magnetism of complex materials, Bose-Einstein condensation, chirality of 
molecules, etc. It was demonstrated there that the profound and innovative idea of quasiaverages formulated by N.N. Bogoliubov, gives the so-called macro-objectivation of the 
degeneracy in domain of quantum statistical mechanics, quantum field theory and in the quantum physics in general.  The complementary unifying 
ideas of modern physics, namely: spontaneous symmetry breaking, quantum protectorate and emergence were discussed also.\\ 
The interrelation of the 
concepts of symmetry breaking, quasiaverages and quantum protectorate was analyzed in the context of quantum theory and statistical physics. The 
leading idea was the statement of F. Wilczek~\cite{wil05}: "primary goal of fundamental physics is to discover profound concepts that 
illuminate our understanding of nature". The works of N.N. Bogoliubov on microscopic theory of superfluidity and superconductivity as well as on 
quasiaverages and broken symmetry belong to this class of ideas. Bogoliubov's notion of quasiaverage  is an essential conceptual advance of 
modern physics, as well as the later concepts of quantum protectorate and emergence. These concepts manifest the operational ability of the 
notion of symmetry; they also demonstrate the power of the unification of various complicated phenomena and have  certain predictive ability. 
Broadly speaking, these concepts are unifying and profound ideas "that illuminate our understanding of nature". In particular, Bogoliubov's 
method of quasiaverages   gives the deep foundation and clarification of the concept of broken symmetry. It makes the emphasis on the notion of   
degeneracy and plays an important role in equilibrium statistical mechanics of many-particle systems. According to that concept, infinitely small 
perturbations can trigger macroscopic responses in the system if they break some symmetry and remove the related degeneracy (or quasi-degeneracy) 
of the equilibrium state. As a result, they can produce macroscopic effects even when the perturbation magnitude is tend to zero, provided that 
happens after passing to the thermodynamic limit. This approach has penetrated, directly or indirectly, many areas of the contemporary physics. 
Practical techniques covered include quasiaverages, Bogoliubov theorem on the singularity of $1/q^2$, Bogoliubov's inequality, and its applications 
to condensed matter physics.\\ 
Condensed matter physics is the field of physics that deals with the macroscopic physical properties of matter. In 
particular, it is concerned with the condensed phases that appear whenever the number of constituents in a system is extremely large and the 
interactions between the constituents are strong. The most familiar examples of condensed phases are solids and liquids. More exotic condensed 
phases include the superfluid and the Bose-Einstein condensate found in certain atomic systems. In condensed matter physics, the symmetry is 
important in classifying different phases and   understanding the phase transitions between them. The phase transition is a physical phenomenon 
that occurs in macroscopic systems and consists in the following. In certain equilibrium states of the system an arbitrary small influence leads 
to a sudden change of its properties: the system passes from one homogeneous phase to another. Mathematically, a phase transition is treated as a 
sudden change of the structure and properties of the Gibbs distributions describing the equilibrium states of the system, for arbitrary small 
changes of the parameters determining the equilibrium~\cite{minl}. The crucial concept here is the order parameter. In statistical physics the question of 
interest is to understand how the order of phase transition in a system of many identical interacting subsystems depends on the degeneracies of 
the states of each subsystem and on the interaction between subsystems. In particular, it is important to investigate a role of the symmetry and 
uniformity of the degeneracy and the symmetry of the interaction. Statistical mechanical theories of the system composed of many interacting 
identical subsystems have been developed frequently for the case of ferro- or antiferromagnetic spin system, in which the phase transition is 
usually found to be one of second order unless it is accompanied with such an additional effect as spin-phonon interaction. Second order phase 
transitions are frequently, if not always, associated with spontaneous breakdown of a global symmetry. It is then possible to find a 
corresponding order parameter which vanishes in the disordered phase and is nonzero in the ordered phase. Qualitatively the transition is 
understood as condensation of the broken symmetry charge carriers. The critical region is reasonably described by a local Lagrangian involving 
the order parameter field. Combining many elementary particles into a single interacting system may result in collective behavior that 
qualitatively differs from the properties allowed by the physical theory governing the individual building blocks. This is the essence of the
emergence phenomenon.\\  
It is known that the 
description of spontaneous symmetry breaking that underlies the connection between classically ordered objects in the thermodynamic limit and 
their individual quantum-mechanical building blocks is one of the cornerstones of modern condensed-matter theory and has found applications in 
many different areas of physics. The theory of spontaneous symmetry breaking, however, is inherently an equilibrium theory, which does not 
address the dynamics of quantum systems in the thermodynamic limit. Any state of matter is classified according to its order, and the type of 
order that a physical system can possess is profoundly affected by its dimensionality. Conventional long-range order, as in a ferromagnet or a 
crystal, is common in three-dimensional systems at low temperature. However, in two-dimensional systems with a continuous symmetry, true 
long-range order is destroyed by thermal fluctuations at any finite temperature. Consequently, for the case of identical bosons, a uniform 
two-dimensional fluid cannot undergo Bose-Einstein condensation, in contrast to the three-dimensional case. 
The two-dimensional system 
can be effectively investigated on the basis of Bogoliubov' inequality. Generally inter-particle interaction is responsible for 
a phase transition. But Bose-Einstein condensation type of phase transition occurs entirely due to the Bose-Einstein   statistics. The typical 
situation is a many-body system made of identical bosons, e.g. atoms carrying an integer total angular momentum. To proceed one must construct 
the ground state. The simplest possibility to do so occurs when bosons are non-interacting. In this case, the ground state is simply obtained by 
putting all bosons in the lowest energy single particle state, as the brilliant Bogoliubov's theory describes.\\ 
From the other hand, it is clear that only a thorough experimental and theoretical 
investigation of quasiparticle many-body dynamics of the many-particle systems can provide the answer on the relevant 
microscopic picture~\cite{kuz09}. As 
is well known, Bogoliubov was first to emphasize the importance of the time scales in the many-particle systems thus anticipating the concept of 
emergence of macroscopic irreversible behavior starting from the reversible dynamic equations.   More recently it has been possible to go step 
further. This step leads to a deeper understanding of the relations between microscopic dynamics and macroscopic behavior on the basis of 
emergence concept~\cite{rbl}. Emergence - macro-level effect from micro-level causes - is an important and profound interdisciplinary notion of modern 
science. There has been renewed interest in emergence within discussions of the behavior of complex systems. It is worth also noticing that the 
notion of quantum protectorate~\cite{rbl} complements the concepts of broken symmetry and quasiaverages by making emphasis on the hierarchy of the 
energy scales of many-particle systems. In an indirect way these aspects arose already when considering the scale invariance and spontaneous 
symmetry breaking. D.N. Zubarev showed~\cite{zub} that the concepts of symmetry breaking perturbations and quasiaverages play an important role in the 
theory of irreversible processes as well. The method of the construction of the nonequilibrium statistical operator becomes especially deep and 
transparent when it is applied in the framework of the quasiaverage concept. For detailed discussion of the Bogoliubov's ideas and methods in the 
fields of nonlinear oscillations and nonequilibrium statistical mechanics see Refs.~\cite{nnb1,kuz07,nnbj}. 
It was demonstrated in Ref.~\cite{kuz10} that the connection and interrelation of the conceptual advances of the many-body physics discussed above show 
that those concepts, though different in details, have complementary character. Many problems in the field of  statistical physics of  complex 
materials and systems (e.g. the chirality of molecules) and the foundations of the microscopic theory of magnetism and superconductivity were 
discussed in relation to these ideas.\\ 
To summarize, it was demonstrated that the Bogoliubov's method of quasiaverages plays a fundamental role in equilibrium and nonequilibrium 
statistical mechanics and quantum field theory and is one of the pillars of modern physics. It will serve for the future development of physics 
as invaluable tool.   All the methods developed by N. N. Bogoliubov are and will remain the important core of a theoretician's toolbox, and of 
the ideological basis behind this development.  Additional material and discussion of these problems can be found in   
recent publications~\cite{kuz09b,kuz10a}.

\end{document}